\documentclass[epj,nopacs,draft]{svjour}

\usepackage{amsfonts,amssymb,amsbsy}
\usepackage[mathscr]{eucal}
\newcommand{\ud}{\mathrm{d}}

\begin{document}



\title{Decoupling of Higgs boson from the inflationary stage of Universe
evolution}

\author{V.V.Kiselev\inst{1,2}, S.A.Timofeev\inst{1,2}}
\institute
{Russian State Research
Center ``Institute for High
Energy Physics'', 
Pobeda 1, Protvino, Moscow Region, 142281, Russia,
\email{Valery.Kiselev@ihep.ru} \and
{Moscow Institute of Physics and
Technology, Institutskii per. 9, Dolgoprudnyi, Moscow Region, 141701,
Russia} }

\abstract{
The constraint on the mass of Higgs field in the Standard Model at the
minimal interaction with the gravity is derived in the form of lower
bound $m_H> 150$ GeV by the strict requirement of decoupling the Higgs
boson from the inflation of early Universe: the inflation produced by
the Higgs scalar could crucially destroy visible properties of large
scale structure of Universe, while the large mass makes the Higgs
particle \textit{not able} to produce the inflation and shifts its
cosmological role into the region of quantum gravity.}
\date{}
\maketitle



\section{Introduction} At present, in cosmology the inflation stage has
became the commonly recognized model for the evolution of early
Universe \cite{infl,i-Linde,i-Albrecht+Steinhardt,i-Linde2,inflation}.
In the simplest scenario consistent with the observed anisotropy of
cosmic microwave background radiation (CMBR) \cite{wmap,wmap-prime},
the supernova data \cite{Astier,Riess,WoodVasey} and large-scale
structure of Universe (LSS) \cite{inhom}, a scalar field of inflaton
should possess some specific properties: an almost flat potential of
self-action with the energy density of the order of
$(10^{16}\mbox{GeV})^4$.

In this respect, recently the possibility of producing the inflation
stage by the Higgs boson with a non-minimal coupling to the gravitation
has been studied
\cite{Barvinsky,Bezrukov,DeSimone,Barvinsky2,Bezrukov2}. Then, in
addition to the Einstein--Hilbert Lagrangian of gravitational field
$$L_\mathrm{EH}=-\frac{1}{16\pi G}R,$$ the interaction Lagrangian has
included the term of the form ${\xi {\Phi}^\dagger {\Phi} R}$, where
$\Phi$ denotes the Higgs field, $R$ is the Ricci scalar, $G$ is the
gravitational constant, and $\xi$ is the coupling constant. So, an
appropriate conformal transformation introduces an effective field
minimally coupled to the gravity, while a relevant effective potential
has got a flat plateau with an altitude regulated by the parameter
$\xi$. The realistic model suggests $\xi\gg 1$, and in this respect, a
new dynamical scale is introduced by $M_\mathrm{Pl}/\xi$
\cite{Burgess,Barbon}, where the Planck mass $M_\mathrm{Pl}$ is defined
by the gravitational constant $G$ as $M_\mathrm{Pl}^2=1/G$. The new
scale determines the altitude of potential plateau and produces the
threshold for changing a regime of coupling at ultraviolet
virtualities. In this way, some definite constraints on the mass of
Higgs boson have been derived. A lower bound for the Higgs boson mass
is determined by the cosmology data (mainly, a slope of primary
spectrum for the inhomogeneity of matter distribution), while the upper
bound is set by a reasonable weakness of the self-coupling.

Similarly, a modified model of induced gravity with the Higgs boson
non-minimally coupled to the gravity was considered in
\cite{CervantesD,CervantesD-GUT}. In that approach, the Einstein--Hilbert term
with the bare gravitational constant is excluded from the primary action.
This formulation leads to an essential change of cosmological dynamics due to a
varying gravitational constant. In this way, the inflationary evolution
in the model results in the fact that the observed contrast of energy
density in the Universe gives the strict preference for extremely large
values of Higgs boson mass though below the Planck mass by four or five
orders of magnitude.

As for the Higgs field nonminimally coupled to the gravity, it can successfully
produce the inflation compatible with the observed properties of Universe.
However, in that case one should introduce the additional coupling constant of
Higgs field with the gravity at a specific value of such the parameter. Then,
the mechanism of inflation caused by the Higgs field nonminimally coupled to the
gravity essentially differs from the mechanism with the Higgs field minimally
coupled to the gravity, so that we consider the minimal version in detail with
no further reference to the nonminimal version beyond the Introduction.

In this paper we consider conditions of developing the inflation stage
produced by the Higgs field of Standard Model (SM) minimally coupled to
the gravity\footnote{In principle, the coupling $\xi$ could deviate from
zero by a small value, which can be neglected in the consideration, say, at
$|\xi|\ll 1$.}, i.e. in the case of $\xi=0$, when the renormalizable
potential of Higgs scalar is determined by two parameters: a vacuum
expectation value fixed by the Fermi constant $G_F$ in the weak
interaction, and a mass not yet measured experimentally. In theoretical
models treating the inflation of Universe due to the scalar field, the
mass of inflaton should get a value at the scale of ${10^{13}}$~GeV in
order to agree with the observed spectra of inhomogeneities in both
CMBR and LSS. In contrast, the mass of Higgs boson is not greater than
several hundreds GeV as follows from SM  with account of loop
corrections including the Higgs boson (the current status of Higgs
particle physics see in review \cite{Quigg}, while the probable fate of
SM is discussed in \cite{JEllis}).  In addition, the observations particularly
require an extremely small constant of self-action for the scalar field
producing the fluctuations transformed into the inhomogeneity of matter and
anisotrophy of cosmic microwave background radiation. So, we will use this
argumentation in our motivation to theoretically forbid the inflation produced
by the Higgs field minimally coupled to the gravity.

Thus, the main conclusion could be
drawn as follows: the scenario, when the Higgs boson alone generates the
Universe inflation consistent with the modern cosmological observations, is
\textit{experimentally forbidden}.
However, the Higgs boson is an ordinary scalar field, hence, in the
framework of classical gravitation theory, the field is \textit{able}
to produce the inflation regime. In practice, no traces of such the
specified inflation have been yet observed. Then, we have to conclude
that in the very beginning at a sufficiently high density of energy
there could be, at least, two options in producing the inflation of early
Universe either by the standard Higgs boson or by the special scalar,
and the second scenario with the special field actually occurred\footnote{In
the former case, one could imagine the scenario, when the inflation caused by
Higgs boson at high density of energy, is further transformed to the regime of
inflation driven by the scecific inflaton with the more flat potential at lower
density of energy. Then, one should suggest a fine tunning, since the change
of regime would be before the end of inflation generated by the Higgs boson,
otherwise one meets with the same experimental constraints mentioned above.}.
What was a reason for such the discrimination between two possible ways of
evolution? We argue for the situation when the inflation originated by
the special inflaton field had got no alternative \textit{if the mass
of Higgs scalar in SM exceeds a critical value}, so that \textit{the
inflation generated by the Higgs boson could not develop in principle.} Such
the critical value of Higgs boson mass defines the decoupling of Higgs boson
from the inflation, at all.

Definitely, in the framework of quantum field theory in the curved
spacetime, when the gravity is treated as a classical theory, there is
a critical value of the Hubble constant, which sets the end of
inflation regime depending on the parameters of inflaton. At lower
values of the Hubble constant, a transition to a reheating of the
Universe occurs due to generating various quanta of both the inflaton
as well as other matter fields. The stage of reheating is properly
called the moment of ``Big Bang''. However, the critical value of
Hubble constant and, hence, the corresponding energy density could be
quite great, so that the gravity would not allow the classical
description, i.e. quantum fluctuations in a metric would be essential,
and the theory enters the scope beyond the validity of inflationary
theory. Thus, a border of quantum gravity in cosmology could actually
determine the decoupling of Higgs boson from the inflation regime.

Essential quantum fluctuations should be inevitably introduced, if the
classical gravitational action $S$ with account of term due to the
relevant inflaton potential becomes comparable with the period of quatum
mechanical amplitude $\Psi$ taken in the classical limit,
$\Psi\sim\exp\{\mathrm{i}S\}$. 
The action in cosmology is related to the Hubble rate $H$ for the Universe
expansion, hence, the curvature of space-time. The curvature of Planckian scale
is beyond the classical description\footnote{This fact was originally
recognized in \cite{i-Linde2}, where A.Linde used it to set the constraint on
the scalar field self-coupling constant $\lambda$ by the
order of magnitude, as was recently rederived in \cite{GK}, $\lambda\ll
10^{-2}$. In the present paper we get an exact value for the critical
value of self-coupling constant, but the order of magnitude result.}. However,
it turns out formally, that the
Higgs boson with a rather large mass would classicaly produce the inflation at
the Planckian curvature of space-time, i.e. at the stage, when the quantum
description of gravity cannot be ignored, hence, the inflationary regime cannot
be induced.

In the present paper, we estimate a lower bound for the Higgs boson mass by
requirement that the Higgs field cannot produce the inflation regime at
early stages of the Universe evolution. The decoupling mass of the
Higgs particle is quite actual for modern experimental searches of
Higgs boson at colliders \cite{Quigg,JEllis}.

Some other aspects of Higgs particle physics as concerns for the
inflation, basically for various fluctuations, were considered in
\cite{Espinosa}.

We have tried to treat rather a complex problem to distinguish between two fine
possibilities, when
\begin{enumerate}
 \item 
the scalar field is able to produce the inflation of universe, but the
parameters of such the inflation would be in a sharp conflict with the observed
properties of our Universe, and therefore, this fact is leading to the
conclusion that such the inflation should be \textit{forbidden experimentally};
\item
the scalar field is not able to produce the inflation of universe, since such
the inflation is \textit{forbidden theoretically} by some critical properties of
field
self-action.
\end{enumerate}

	The first of above possibilities occurs if the value of self-action
coupling $\lambda$ for the Higgs particle minimally coupled to the gravity is
below the critical value\footnote{Extremely small values of constant
$\lambda$, when the inflation generated by the Higgs boson further develops due
to switching into the regime driven by the specific inflaton, are excluded
experimentally: $\lambda>0.11$ (seediscussion in \cite{GK}).}, while the second
possibility occurs if the coupling $\lambda$ exceeds the critical value. 
	Thus, we can discriminate two answers to the question: why the observed
Universe did not evolve at early times through the inflationary stage produced
by the scalar Higgs particle of Standard model? The first answer is the
following: the inflation produced by the Higgs field was occasionally missed,
since the other scalar field with a specific properties has produced the
different inflation. The second answer states that the inflation produced by the
Higgs field cannot exist because the Higgs particle is too heavy to produce the
inflation. The second statement is determinative, while the first answer leads
to the problem of preference for one of two scenarios of inflation by  the
specific inflaton or Higgs particle giving very different post-inflationary
universes.

\section{The Higgs boson as the inflaton} Let us consider the model of
the Higgs boson in the gauge setting the real field
${\Phi}=\phi/\sqrt{2}$ with the minimal coupling to the gravity and
potential\footnote{We assume, that a nonzero cosmological constant can
be surely neglected during the inflation.} $$V=\lambda(\phi^2-v^2)^2/4,$$
where the vacuum expectation value $v=1/\sqrt{\sqrt{2}G_F}\approx
246.2\mbox{ GeV}$ is known experimentally. Then, the formula for the
mass of Higgs field is ordinary given by
\begin{equation}
m^2=2\lambda v^2.
\end{equation}

The Einstein--Hilbert action of gravity is classically defined by
scalar curvature $R$
\begin{equation}
S_g=-\frac{1}{16\pi G}\int \ud^4x \sqrt{-g}\,R,
\end{equation}
while the cosmology in the case of spatial homogeneity is described by
the metric with a time-dependent scale factor $a(t)$,
\begin{equation}\label{scale-fact}
    \ud s^2=\ud t^2-a^2(t)\,\ud \boldsymbol r^2.
\end{equation}
In the inflation regime, the metric can be well approximated by de
Sitter one in the standard cosmological form, wherein the 3-dimensional
space is flat and an observer is posed in the center point
\begin{equation}\label{dS}
    \ud s^2 \approx \ud t^2 - e^{2Ht}\,\ud \boldsymbol r^2,
\end{equation}
where $H=\mathrm{d}\ln a/\mathrm{d}t=\dot a/a$ is the Hubble constant, which
value to the end of inflation can be \textit{strictly} related with the
constant $\lambda$ defining the self-coupling for the Higgs field.

The appropriate framework of quasiattractor approach is systematically
simple: the motion can be straightforwardly treated in terms of
autonomous differential equations with a parametric attractor, whose
critical points slowly drift with the Hubble constant\footnote{The
scale factor runs as the exponent of e-folding $N$ by definition
$a\sim\exp\{- N\}$, while the Hubble constant gets a slow driftage logarithmic
in the scale factor, more exactly, linear in e-folding for the quartic
self-action of inflaton, $H-H_*\sim N$.} \cite{Mexicans,KT3,KT-GERG}.

Indeed, in terms of dimensionless variables defined as
\begin{equation}\label{evol-3}
    x=\frac{\kappa}{\sqrt{6}}\,\frac{\dot \phi}{H}, \qquad
    y=\sqrt[4]{\frac{\lambda}{12}}\frac{\kappa\,\phi}{\sqrt{\kappa
    H}},\qquad
    z=\frac{\sqrt[4]{3\lambda}}{\sqrt{\kappa H}},
\end{equation}
giving the fractions of kinetic energy $x^2$ and potential energy $y^4$ for
the energy budget of scalar field: $x^2+y^4=1$ at $\kappa^2=8\pi G$,
while $z$ introduces the parametric dependence in the field equations
\begin{equation}\label{evol-6}
    x^\prime=
    -3x^3+3x+2y^3z,\qquad
    y^\prime=
    -\frac{3}{2}\,x^2y-xz,
\end{equation}
wherein the evolution, i.e. the differentiation denoted by prime, is calculated
with respect to e-folding defined by $N=\ln
a_\mathrm{end}-\ln a$, so that the parameter $z$ evolves according to
\begin{equation}\label{parameter-z}
    z^\prime=
    -\frac{3}{2}\,x^2z.
\end{equation}
There are stable critical points\footnote{Critical points are defined
by condition $x^\prime=y^\prime=0$, while the stability takes place
when linear perturbations in differential equations near the critical
points decline to zero.} for (\ref{evol-6}) (see \cite{KT3,KT-GERG}) at
\begin{equation}\label{z-stab}
    z^4<\frac{3}{4}.
\end{equation}

The existence of critical points is caused by specific ``friction
term'' in equations. The magnitude of friction is given by the Hubble
constant, so that the kinetic energy of inflaton is suppressed with
respect to the potential, and the field slowly rolls down to the
minimum of potential. At large amount of e-folding $N\gg 1$, the
quasiattractor is equivalent to the slow-roll approximation in the
leading order of $1/N$-expansion \cite{deVegas}. However, in contrast
to the slow-rolling in the $1/N$-expansion, the quasiattractor allows
us to get the \textit{strict} description for the final stage of inflation due
to the \textit{exact} determination of critical points for the parametric
attractor. Reasonably, the inflation finishes at such value of Hubble
rate, when a condition on the existence of stable critical points
invalidates and the attractor becomes unstable, i.e. it disappears.
Then, from the condition of (\ref{z-stab}) we get
\begin{equation}\label{five}
{2\pi}G H^2_\mathrm{end}=\lambda.
\end{equation}
To the end of inflation evolving with the parametric attractor we get
$z^4_\mathrm{end}=3/4$, $x^2_\mathrm{end}=2/3$ and
$y^4_\mathrm{end}=1/3$, hence, $H^2_\mathrm{end}=2\pi G
\lambda\phi^4_\mathrm{end}$, so that equivalently to (\ref{five}) we
get
\begin{equation}\label{e-five}
    2\pi G\phi^2_\mathrm{end}=1.
\end{equation}
Thus, the inflation produced by the Higgs boson stops at the Planckian
scale of field value.

From (\ref{five}) we see that if $\lambda\sim 1$, the value of $H_\mathrm{end}$
is about the Planck scale. This result repeats the arguments of
\cite{i-Linde2,GK} as mentioned above. 
Therefore, the heavy Higgs boson formally corresponds
to the inflationary Hubble rate about the Planckian scale of energy, where
effects of  quantum gravity cannot be ignored, hence, the inflation dynamics
cannot develop, since the curvature of space-time gets the Planckian values.

\section{A border of quantum gravity in cosmology} De Sitter metric
(\ref{dS}) straightforwardly determines the scalar curvature standing
in the action of classical gravitational field, $R=-12H^2$.

In the calculation of gravitational action, it is worths to note, that
the coordinate $r$ takes values in the region from zero to the horizon
$r_H=1/H$ and the integration in time $t$ is limited by the interval
from the negative infinity to a moment, which can be put to zero with
no lose of generality of consideration (in fact, to a moment of
inflation end). Notice, that the specified coordinate system covers
only a half of de Sitter manifold\footnote{The other half of manifold
can be associated with the exponentially contracting universe, in
contrast to the case of expanding universe, we consider.}, therefore,
the action can be doubled, in principle, but this would incorporate a
part of the manifold causally independent of the cosmological
observer. 
Finally, we get
\begin{equation}
S_g=\frac{1}{3GH^2}.
\end{equation}

Similarly, we add the contribution of action for the matter
approximated by the form
\begin{equation}
S_m\approx -\int \ud^4 x \sqrt{-g}\, V,
\end{equation}
where $V$ is the matter potential, so that in the framework of
inflation regime we neglect the kinetic term of inflaton in comparison
with the potential. The contribution of potential is determined by the
Einstein equations
$$V\approx\frac{3H^2}{8\pi G}.$$

Finally, the matter action equals
\begin{equation}
S_m=-\frac{1}{6GH^2},
\end{equation}
yielding the sum $S=S_g+S_m$ equal to
\begin{equation}\label{action-sum}
    S=\frac{1}{6GH^2}.
\end{equation}

We have just got the action by making use of de Sitter metric. For the
sake of generality, we have performed exact calculations in the case of
matter with a state parameter $w$ equal to the ratio of pressure $p$ to
energy density $\rho$: $p=w\rho$.  In the range of $-1<w<1$, the
integration in time runs along a finite interval with the scale factor
spanning the region from a cosmic singularity to the moment defined by
$a=1$. Then, \textit{the action of matter and gravity takes the same
value of} {(\ref{action-sum})} \textit{independent of the state
parameter} $w$. This fact is important, since it points to the
stability of cosmological action versus the matter content. In
addition, the spacetime to the end of inflation produced by the Higgs
boson becomes to essentially differ from de Sitter spacetime: the
fractions of kinetic and potential energies get values equal to
$\frac{2}{3}$ and $\frac{1}{3}$, correspondingly, that gives
$w=\frac{1}{3}$ specific for a radiation, i.e. a light-like or
ultra-relativistic matter. Nevertheless, the estimate of
(\ref{action-sum}) is rather universal, it does not significantly vary
with changes in the expansion regime.

The quantum mechanical amplitude $\Psi$ in the classical limit takes the
form of $\Psi\sim \exp\{\mathrm{i}S\}$, therefore, in order to separate
the quantum regime from the classical behavior, one should compare the
action with the period $\delta S=2\pi$.
Then, we get the constraint on $H$, when the quantum gravity effects
become essential
\begin{equation}
12\pi GH^2> 1.
\end{equation}
The confidence level of such the constraint is discussed in section
\ref{confidence}.

\section{The mass of decoupling the Higgs boson from the inflation} For
the case of inflation produced by the Higgs scalar, the relation of
Hubble constant at the end of inflation with the self-action constant
$\lambda$ results in
\begin{equation}\label{l-crit}
\lambda>\frac{1}{6}.
\end{equation}
Thus, the constraint on the Higgs mass takes the form
\begin{equation}\label{f}
m > \frac{v}{\sqrt{3}}.
\end{equation}

Substituting the experimental inputs,
we get the lower bound for the Higgs mass as
$m_\mathrm{min}=142.3$~GeV.

\section{The confidence level  of decoupling constraint\label{confidence}}

The decoupling mass obtained from the theoretical forbidding the inflation
produced by the Higgs boson is based on the breaking the classical description
due to quantum fluctuations, which make the inflation impossible. So, in order
to estimate the confidence level of such the lower bound derived above, let us
consider, for instance, a harmonic oscillator with the Hamiltonian
\begin{equation}
      H=\frac12 (Q^2+P^2)\,\hbar\omega,
\end{equation} 
wherein $Q$ is the coordinate, while $P$ is its canonically conjugated momentum.
The state maximally close to the classical system with the energy
$E=\hbar\omega(n+\frac12)$ is  the coherent state with the minimized fluctuation
of coordinate and momentum at any time of evolution
\begin{equation}
      (\delta Q)^2_\mathrm{c}=(\delta P)^2_\mathrm{c}=\frac12,
\end{equation} 
while the stationary quantum state with the definite energy of $n$ quanta gives
essential
fluctuations
\begin{equation}
      (\delta Q)^2_\mathrm{q}=(\delta P)^2_\mathrm{q}=n+\frac{1}{2}.
\end{equation} 
Therefore, one could estimate the relevance of the state assignement to the
nonclassical system by evaluating 
\begin{equation}\label{chi1} 
      \chi^2=\frac{(\delta Q)^2_\mathrm{q}}{(\delta
      Q)^2_\mathrm{c}}-\chi^2_0=2n,
\end{equation} 
wherein we put $\chi^2_0=1$ in order to match the vacuum to completely the 
quantum state.

The same criterium can be derived by considering the time evolution of average
coordinate in the coherent state. So, from
\begin{equation}
      Q(t)=Q_0\cos \omega t+P_0\sin \omega t,
\end{equation} 
one gets
\begin{equation}
      \langle Q(t)\rangle=0,\qquad
      \langle Q^2(t)\rangle=\frac12 (Q_0^2+P_0^2)=n,
\end{equation} 
so that the fluctuation is equal to
\begin{equation}
      (\delta Q)^2_\mathrm{t}=n,
\end{equation} 
that yields
\begin{equation}\label{chi2} 
      \chi^2=\frac{(\delta Q)^2_\mathrm{t}}{(\delta Q)^2_\mathrm{c}}=2n.
\end{equation} 
Then, we can estimate the hypothesis that the system essentially requires
to carefully take into account important quantum fluctuation by the
$\chi^2$-probability depending on the number of quanta in the system. The
evaluation of $n$ is system-dependent. We put
\begin{equation}\label{nq} 
 n=\frac{S}{2\pi},
\end{equation} 
wherein $S$ is the action of the cosmological system with the inflaton. The
lower bound $m_H>m_\mathrm{min}$ is equivalent to $n<1$ and, hence, $\chi^2<2$,
so that the system is essentially quantum within the $2\sigma$ confidence level,
i.e. with the probability of 90\%. 

Note, that changing the determination for the number of quanta in (\ref{nq})  by
$n=S/\pi$ or $n=S/4\pi$ would result in the respective modification of
confidence level for the lower cosmological bound for the Higgs boson mass: 99\%
or 68\%, correspondingly. Analogously, the shift $m_\mathrm{min}\mapsto \sqrt{2}
m_\mathrm{min}$ would dicrease the confidence level of such the estimate with
(\ref{nq}) to the value of 68\%.

\section{A renormalization group improvement} The above consideration
has been based on the leading approximation of effective action, while
quantum loop corrections would both modify the potential at large
fields relevant to the inflation and renormalize the physical
parameters of Lagrangian for the Higgs scalar, i.e. the filed
normalization, mass and coupling constant $\lambda$. These effects
could be effectively taken into account by making use of
renormalization group in SM \cite{Espinosa} with an appropriate choice
of renormalization point at the inflation stage, so that the whole
effect would be reduced to the running of $\lambda(\mu)$ with the scale
$\mu$. Similar strategy has been explored in
\cite{Bezrukov2,Barvinsky2} for the Higgs field non-minimally coupled
to the gravity. So, in estimates we fit a pole mass of Higgs particle,
determining the running mass $m(\mu)$ at the scale of $t$-quark mass,
with other parameters of SM at the same scale $\mu=m_t$ to reach the
critical value of $\lambda$ in (\ref{l-crit}) at the scale of the order
of the Planck mass. Then, the variation of final result due to the
renormalization group can be estimated by comparing one- and two-loop
calculations, which points to the uncertainty caused by the choice of
final scale in the running. Another source of uncertainties is
connected to the empirical accuracy in the measurement of SM parameters
at the starting scale of renormalization group evolution. So, the
one-loop renormalization group results in the decoupling mass of Higgs
particle equal to 153~GeV, while the two-loop evolution approximately
gives the lower value of 150~GeV at $m_t=171$ GeV. A complete analysis
of uncertainties caused by variation of different parameters in the
calculations by means of renormalization group will be given elsewhere.

Then, the renormalization group improvement of estimate results in the
lower bound for the Higgs boson mass $m_\mathrm{min}\approx 150$ GeV with
uncertainty of 3 GeV. The difference between estimates at the tree
level and due to the two-loop renormalization is significant, but it is
rather moderate, so that we can draw the conclusion that the higher
order corrections are still under control.

\section{Final remarks} It is worth to note, that having estimated the
inflation parameters we have neglected terms quadratic in the Higgs
field, which is correct, if the vacuum expectation value is much less
than the Planck mass, i.e. at $v \ll M_{Pl}$ (this condition is safely
valid for the Higgs boson in SM). Therefore, the estimation of
(\ref{f}) is valid for any scalar Higgs boson with a small vacuum
expectation value in gauge theories including grand unified theories
(GUT). In addition, a grand unification could change both the running
of gauge coupling constants and set of quantum fields active in the
running. Then, the estimate obtained due to the renormalization group
improvement would slightly move, though the value of such displacement
should not sizably exceed the calculation uncertainty given above. In
respect of GUT with the SU(5) symmetry we mention the modified induced
gravity scenario with the Higgs field non-minimally coupled to the
gravity as studied in \cite{CervantesD-GUT}.

\section*{Acknowledgements}
This work was partially supported by grants of Russian Foundations for Basic
Research 09-01-12123 and 10-02-00061, Special Federal Program ``Scientific and
academics personnel'' grant for the Scientific and Educational Center
2009-1.1-125-055-008, ant the work of T.S.A. was supported by the
Russian President grant MK-406.2010.2.

\end{document}